\documentstyle[twoside,epsf]{article}
\catcode`\@=11
\long\def\@makefntext#1{
\protect\noindent \hbox to 3.2pt {\hskip-.9pt  
$^{{\eightrm\@thefnmark}}$\hfil}#1\hfill}		

\def\@makefnmark{\hbox to 0pt{$^{\@thefnmark}$\hss}}	
	
\def\ps@myheadings{\let\@mkboth\@gobbletwo
\def\@oddhead{\hbox{}
\rightmark\hfil\eightrm\thepage}   
\def\@oddfoot{}\def\@evenhead{\eightrm\thepage\hfil
\leftmark\hbox{}}\def\@evenfoot{}
\def\sectionmark##1{}\def\subsectionmark##1{}}

\oddsidemargin=\evensidemargin
\newcounter{sectionc}\newcounter{subsectionc}\newcounter{subsubsectionc}
\renewcommand{\section}[1] {\vspace{12pt}\addtocounter{sectionc}{1} 
\setcounter{subsectionc}{0}\setcounter{subsubsectionc}{0}\noindent 
	{\tenbf\thesectionc. #1}\par\vspace{5pt}}
\renewcommand{\subsection}[1] {\vspace{12pt}\addtocounter{subsectionc}{1} 
	\setcounter{subsubsectionc}{0}\noindent 
	{\bf\thesectionc.\thesubsectionc. {\kern1pt \bfit #1}}\par\vspace{5pt}}
\renewcommand{\subsubsection}[1] {\vspace{12pt}\addtocounter{subsubsectionc}{1}
	\noindent{\tenrm\thesectionc.\thesubsectionc.\thesubsubsectionc.
	{\kern1pt \tenit #1}}\par\vspace{5pt}}

\newcounter{appendixc}
\newcounter{subappendixc}[appendixc]
\newcounter{subsubappendixc}[subappendixc]
\renewcommand{\thesubappendixc}{\Alph{appendixc}.\arabic{subappendixc}}
\renewcommand{\thesubsubappendixc}
	{\Alph{appendixc}.\arabic{subappendixc}.\arabic{subsubappendixc}}

\renewcommand{\appendix}[1] {\vspace{12pt}
        \refstepcounter{appendixc}
        \setcounter{figure}{0}
        \setcounter{table}{0}
        \setcounter{lemma}{0}
        \setcounter{theorem}{0}
        \setcounter{corollary}{0}
        \setcounter{definition}{0}
        \setcounter{equation}{0}
        \renewcommand{\thefigure}{\Alph{appendixc}.\arabic{figure}}
        \renewcommand{\thetable}{\Alph{appendixc}.\arabic{table}}
        \renewcommand{\theappendixc}{\Alph{appendixc}}
        \renewcommand{\thelemma}{\Alph{appendixc}.\arabic{lemma}}
        \renewcommand{\thetheorem}{\Alph{appendixc}.\arabic{theorem}}
        \renewcommand{\thedefinition}{\Alph{appendixc}.\arabic{definition}}
        \renewcommand{\thecorollary}{\Alph{appendixc}.\arabic{corollary}}
        \renewcommand{\theequation}{\Alph{appendixc}.\arabic{equation}}
        \noindent{\tenbf Appendix \theappendixc #1}\par\vspace{5pt}}
\newcommand{\subappendix}[1] {\vspace{12pt}
        \refstepcounter{subappendixc}
        \noindent{\bf Appendix \thesubappendixc. {\kern1pt \bfit #1}}
	\par\vspace{5pt}}
\newcommand{\subsubappendix}[1] {\vspace{12pt}
        \refstepcounter{subsubappendixc}
        \noindent{\rm Appendix \thesubsubappendixc. {\kern1pt \tenit #1}}
	\par\vspace{5pt}}
\newcommand{\textlineskip}{\baselineskip=13pt}
\newcommand{\smalllineskip}{\baselineskip=10pt}

\def\eightcirc{
\begin{picture}(0,0)
\put(4.4,1.8){\circle{6.5}}
\end{picture}}
\def\eightcopyright{\eightcirc\kern2.7pt\hbox{\eightrm c}} 
\newcommand{\copyrightheading}[1]
	{\vspace*{-2.5cm}\smalllineskip{\flushleft
	{\footnotesize International Journal of Theoretical and Applied 
	Finance #1}\\
	{\footnotesize $\eightcopyright$\, World Scientific Publishing
	 Company and Imperial College Press}\\
	 }}

\newcommand{\publisher}[2]{{\begin{center}\footnotesize\smalllineskip 
	#1
	\end{center}
	}}

\def\abstracts#1#2#3{{
	\centering{\begin{minipage}{4.5in}\baselineskip=10pt\footnotesize
	\centerline{\footnotesize ABSTRACT}
	\parindent=0pt #1\par 
	\parindent=15pt #2\par
	\parindent=15pt #3
	\end{minipage}}\par}} 

\def\keywords#1{{
	\centering{\begin{minipage}{4.5in}\baselineskip=10pt\footnotesize
	{\footnotesize\it Keywords}\/: #1
	 \end{minipage}}\par}}
\renewenvironment{thebibliography}[1]
	{\frenchspacing
	 \ninerm\baselineskip=11pt
	 \begin{list}{[\arabic{enumi}]}
	{\usecounter{enumi}\setlength{\parsep}{0pt}
	 \setlength{\leftmargin 13.7pt}{\rightmargin 0pt} 
	 \setlength{\itemsep}{0pt} \settowidth
	{\labelwidth}{[#1]}\sloppy}}{\end{list}}
\newcounter{itemlistc}
\newcounter{romanlistc}
\newcounter{alphlistc}
\newcounter{arabiclistc}

\newcommand{\fcaption}[1]{
        \refstepcounter{figure}
        \setbox\@tempboxa = \hbox{\footnotesize Fig.~\thefigure. #1}
        \ifdim \wd\@tempboxa > 5in
           {\begin{center}
        \parbox{5in}{\footnotesize\smalllineskip Fig.~\thefigure. #1}
            \end{center}}
        \else
             {\begin{center}
             {\footnotesize Fig.~\thefigure. #1}
              \end{center}}
        \fi}
\newcommand{\tcaption}[1]{
        \refstepcounter{table}
        \setbox\@tempboxa = \hbox{\footnotesize Table~\thetable. #1}
        \ifdim \wd\@tempboxa > 5in
           {\begin{center}
        \parbox{5in}{\footnotesize\smalllineskip Table~\thetable. #1}
            \end{center}}
        \else
             {\begin{center}
             {\footnotesize Table~\thetable. #1}
              \end{center}}
        \fi}
\def\@citex[#1]#2{\if@filesw\immediate\write\@auxout
	{\string\citation{#2}}\fi
\def\@citea{}\@cite{\@for\@citeb:=#2\do
	{\@citea\def\@citea{,}\@ifundefined
	{b@\@citeb}{{\bf ?}\@warning
	{Citation `\@citeb' on page \thepage \space undefined}}
	{\csname b@\@citeb\endcsname}}}{#1}}
\newif\if@cghi
\def\cite{\@cghitrue\@ifnextchar [{\@tempswatrue
	\@citex}{\@tempswafalse\@citex[]}}
\def\citelow{\@cghifalse\@ifnextchar [{\@tempswatrue
	\@citex}{\@tempswafalse\@citex[]}}
\def\@cite#1#2{{$\null^{#1}$\if@tempswa\typeout
	{IJCGA warning: optional citation argument 
	ignored: `#2'} \fi}}

\def\pmb#1{\setbox0=\hbox{#1}
	\kern-.025em\copy0\kern-\wd0
	\kern.05em\copy0\kern-\wd0
	\kern-.025em\raise.0433em\box0}

\def\fnt#1#2{\footnotetext{\kern-.3em
	{$^{\mbox{\scriptsize #1}}$}{#2}}}
\def\fpage#1{\begingroup
\voffset=.3in
\thispagestyle{empty}\begin{table}[b]\centerline{\footnotesize #1}
	\end{table}\endgroup}
\def\runninghead#1#2{\pagestyle{myheadings}
\markboth{{\protect\footnotesize\it{\quad #1}}\hfill}
{\hfill{\protect\footnotesize\it{#2\quad}}}}
\font\tenrm=cmr10
\font\tenit=cmti10 
\font\tenbf=cmbx10
\font\bfit=cmbxti10 at 10pt
\font\ninerm=cmr9

\font\eightrm=cmr8

\begin{document}
\runninghead{A Stochastic Cascade Model for FX Dynamics}
{A Stochastic Cascade Model for FX Dynamics}
\normalsize\textlineskip
\thispagestyle{empty}
\setcounter{page}{1}
\copyrightheading{}         
\vspace*{0.88truein}
 \fpage{1}
 \centerline{\bf A STOCHASTIC CASCADE MODEL FOR FX DYNAMICS}
 \vspace*{0.37truein}
 \centerline{\footnotesize WOLFGANG BREYMANN}
 \vspace*{0.015truein}
 \centerline{\footnotesize\it Olsen \& Associates}
 \baselineskip=10pt
 \centerline{\footnotesize\it Seefeldstrasse 233, 8008 Zurich, Switzerland}
 \vspace*{10pt}
 \centerline{\footnotesize SHOALEH GHASHGHAIE}
 \vspace*{0.015truein}
 \centerline{\footnotesize\it Institut f\"ur mathematische Statistik}
 \baselineskip=10pt
 \centerline{\footnotesize\it Sidlerstr. 5, 3012 Bern, Switzerland}
 \vspace*{10pt}
 \centerline{\footnotesize PETER TALKNER}
 \vspace*{0.015truein}
 \centerline{\footnotesize\it Paul Scherrer Institut}
 \baselineskip=10pt
 \centerline{\footnotesize\it 5232 Villigen, Switzerland}
 \vspace*{0.225truein}
 \publisher{}

\vspace*{0.21truein}
 \abstracts {
 A time series model for the FX
dynamics is presented which takes into account structural
peculiarities of the market, namely its heterogeneity and an
information flow from long to short time horizons. The model
emerges from an analogy between FX dynamics and hydrodynamic
turbulence. The heterogeneity of the market is modeled in form of
a multiplicative cascade of time scales ranging from several
minutes to a few months, analogous to the Kolmogorov cascade in
turbulence.
 \newline
The model reproduces well the important empirical characteristics
of FX rates for major currencies, as the heavy-tailed distribution
of returns, their change in shape with increasing time interval,
and the persistence of volatility. }{}{}
\vspace*{5pt}
\section{Information cascade: an analogy with hydrodynamic
  turbulence}
The idea of a heterogeneous market that consists of traders acting
on different time horizons was first advanced by M\"{u}ller et
al.\cite{UAM.1997-01-01} who investigated the absolute values of
FX returns on different time scales. They observed that the price
changes over longer time intervals have a stronger influence on
those over shorter time intervals than conversely. This has been
interpreted as an information flow from long-term to short-term
traders which directly influences the volatility on different time
scales.

Though being very different in its physical nature, the motion of
a turbulent fluid medium is also governed by a hierarchical
process in which energy flows from large to small spatial
scales\cite{uF95}. Interestingly enough, this formal analogy
between FX dynamics and turbulence leads to similar statistical
behavior of both phenomena\cite{SGA.1996-01-01}. For the FX
dynamics, the hierarchical process can be implemented in form of a
volatility cascade, which
has recently been visualized
by means of wavelet techniques\cite{AMS97}.

In the present contribution we take this hypothesis as granted and
construct a stochastic multiplicative volatility cascade driving
the FX price process. In the stochastic cascade model (SCM), that
we present here, the volatility at an instant of time $t$ is
assumed to be created by processes on different time scales
reflecting the heterogeneity of the market dynamics. The
information flow between traders on different time horizons is
modeled by a directional interaction scheme between the levels of
the cascade corresponding to the different time scales. More
precisely, the flow of information is accounted for by an update
of stochastic ``transfer" factors relating the subsequent levels
of the cascade.

\section{Stochastic cascade model}
We denote the price at time $t$ by $p_t$ and define the return as
$r_t \equiv \log p_t - \log p_{t-1}$. In the proposed SCM the
returns are described by a stochastic volatility model, i.e.
\begin{equation}
r_t = \sigma_t \xi_t \label{r},
\end{equation}
where $\xi_t$  are i.i.d. Student random numbers with
3~degrees of freedom.
The volatility $\sigma_t$ is governed by a hierarchical process
that reflects the heterogeneity of the market on its different
time horizons. Each level $k$ of the cascade corresponds to a time
horizon of duration $\tau^{(k)}$. The largest horizon $\tau^{(0)}$
representing the first level at the top of the cascade is
typically of the order of a few months while the level $m$ at the
bottom of the cascade has the smallest horizon $\tau^{(m)}$ on
which dealers may act and which is of the order of
minutes. For the sake of simplicity we assume that the horizon
ratio between neighboring levels is constant, i.e.
$\tau^{(k)}/\tau^{(k-1)} =p <1 $ is independent of $k$.

We now assign to each level $k$ of the hierarchy a volatility
$\sigma^{(k)}_t$ which is determined by the respective volatility
on the level $k-1$ and a time-dependent random factor $a^{(k)}_t$
that will be specified below. Hence the level volatilities are
recursively defined by:
\begin{equation}
\sigma^{(k)}_t = a^{(k)}_t \sigma^{(k-1)}_t. \label{sk}
\end{equation}
As a toy model we assume that the market dynamics is completely
determined by the behavior of the dealers and not subject to
exogenous time-dependent influences. Then, the volatility
$\sigma^{(0)}$ on the largest horizon is time independent. The
volatility on the shortest horizon determines the returns, as
given by eq. (\ref{r}) with $\sigma_t = \sigma^{(m)}_t$. Using eq.
(\ref{sk}) we find:
\begin{equation}
\sigma_t = \sigma^{(0)}\prod_{k=1}^m a^{(k)}_t. \label{s1m}
\end{equation}

\section{Updating rule for the volatility factors $a_t^{(k)}$}

\begin{figure}[t]
\begin{center}
\makebox[1cm]{
 \makebox[14cm][l]{
 \epsfxsize=13cm \epsfbox{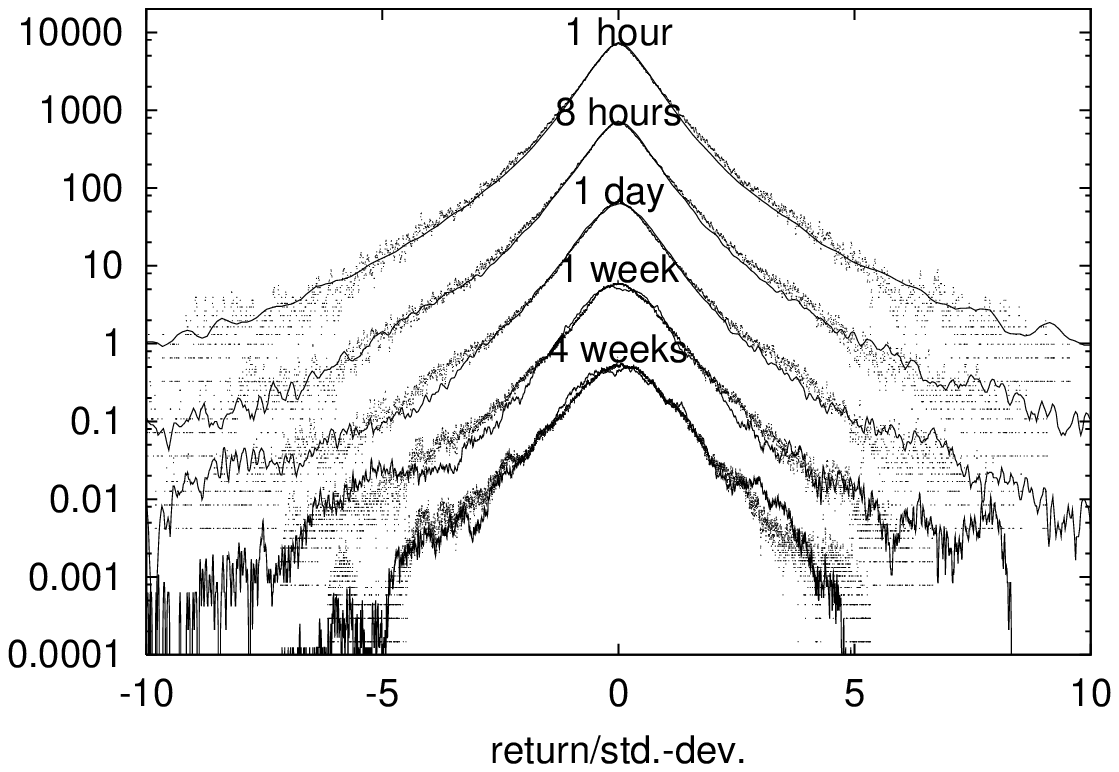}
 }
}
\end{center}
\vspace*{13pt}
  \fcaption{
    Distribution of returns for time intervals reaching from
    1~hour (top) to 4~weeks (bottom). Full lines: SCM simulation
    with parameters $\sigma^{(0)}=0.075$,
    $\tau{(k)}/\tau{(k-1)}=\sqrt{2}$, $\lambda^2_k=0.0173$, and
    $a_k=a$, where $a$ is adjusted
    in such a way that the scaling of the second moment correspond
    to the observed values.
    Dots: USD/CHF FX spot rates from 1986 till 1996.
    For better visibility, the curves have been vertically shifted
    with respect to each other by
    multiples of $\log 10$.
    \label{fRetPdf}
  }
\end{figure}

The time-dependence of the factors $a_t^{(k)}$ results from the
following renewal scheme: At the initial time $t_0$ the factors
$a^{(k)}_{t_0}$ are drawn from independent lognormal distributions
$LN(a_k , \lambda_k^2)$ with mean values $a_k$ and variances
$\lambda_k^2$, $k=1, \ldots, m$. At a later time $t_{n+1}$ the
factor $a^{(1)}_{t_{n+1}}$ at the top of the hierarchy maintains
the corresponding value $a^{(1)}_{t_n}$ at the preceding time
$t_n$ with a probability $1-w^{(1)}$ and else is drawn from
$LN(a_1,\lambda^2_1)$. In the latter case all the subsequent
factors $a_{t_{n+1}}^{(k)}$, $k>1$ are also renewed, i.e. they
aquire independent random values from the respective distributions
$LN(a_k, \lambda^2_k)$. If $a^{(1)}_{t_{n+1}}$ coincides with
$a^{(1)}_{t_n}$ the factor $a_{t_{n+1}}^{(2)}$ will be renewed
with a probability $w^{(2)}$. These rules apply through the whole
cascade down to $k=m$: A renewal at some level~$k$ entails one at
all higher levels $k'>k$; if no renewal has taken place up to
level $k$, the coefficient at level $k+1$ will be renewed with
probability $w^{(k+1)}$ (drawn from the distribution $LN(a_{k+1}
\lambda^2_{k+1})$) or, with probability $1-w^{(k+1)}$, its value
remains constant in time, $a^{(k+1)}_{t_{n+1}} = a^{(k+1)}_{t_n}$.

The renewal probabilities $w^{(k)}$ are given by:
\begin{equation}
w^{(k)} = 1-(1-p^{m-k})/(1-p^{m-k+1}), \quad k=1,\ldots, m.  \label{wk}
\end{equation}
According to this renewal scheme the mean life-times of a factor
at level k (measured in no. of time steps of length $\tau~{(m)}$)
is given by $p^{k-m}$ where $p$ is the scaling factor of the time
horizons as defined above.

\section{Results}

 \begin{figure}[tbh]
 \begin{center} \makebox[1cm]{
 \makebox[14cm][l]{
 \epsfxsize=13cm \epsfbox{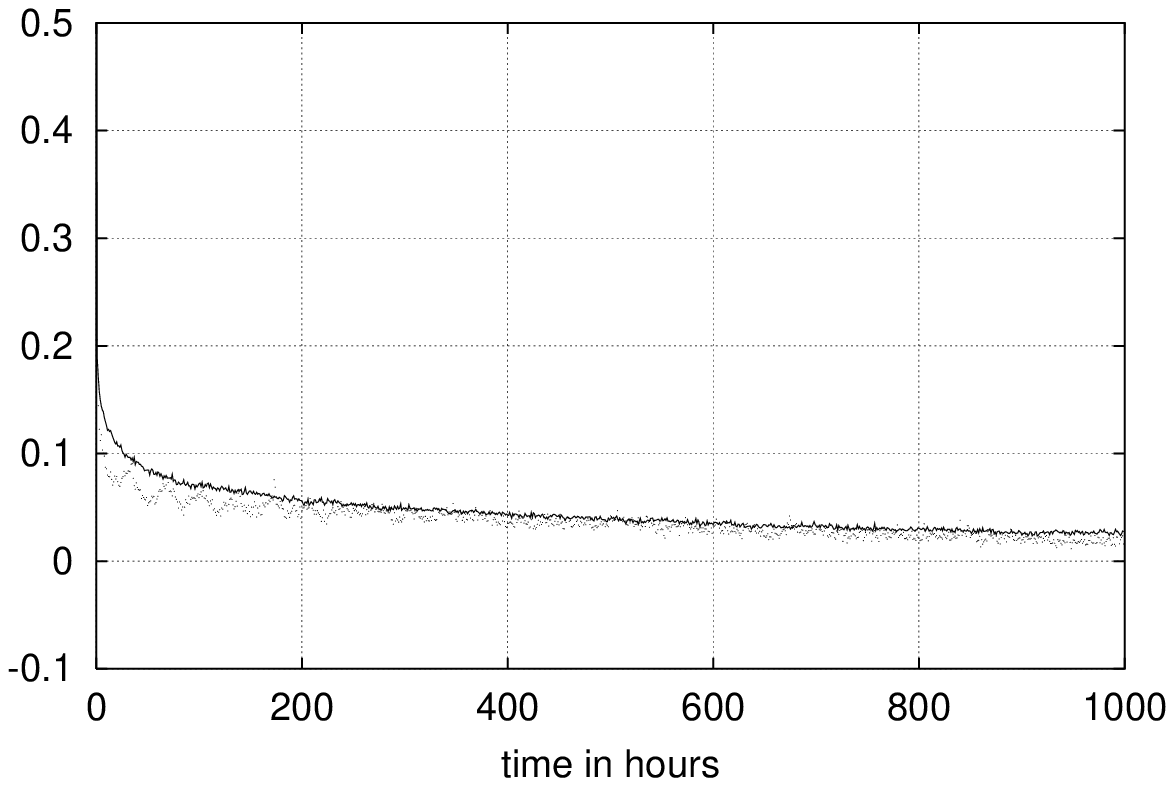}
 }
}
\vspace*{13pt}
  \fcaption{\label{fVolatilityAC}
    \footnotesize \sf
Autocorrelation function of absolute returns for the same
simulation (full lines) and observed data (dots) as in Fig. 1.
  }
\end{center}
\end{figure}

The time series defined by  eqs. (\ref{r} - \ref{wk}) have been
simulated and the parameters have been adjusted in such a way that
the model simultaneously reproduces the empirical return
distributions (Fig. \ref{fRetPdf}), the scaling laws for the
moments\cite{BGTtbp}, and the autocorrelation function of the
volatility (Fig. \ref{fVolatilityAC}) of USD/CHF FX spot rates.
The distributions of returns (cf. Fig. \ref{fRetPdf}) exhibit the
well-known heavy tails, which lose weight with increasing time
interval. The autocorrelation function of the absolute returns
(Fig. \ref{fVolatilityAC}) shows the characteristic slowly
decaying tail. In both cases the SCM simulation (full line)
reproduces the observed data (dotted curves) very well. The
scaling behavior of the moments and the cross correlation function
of the volatility are presented in an extended version of the
paper\cite{BGTtbp}.

To summarize, the SCM is a hierarchical time series model where a net
information flow from long to short time horizons is implemented
in terms of a random unidirectional action of the volatility at a
given time horizon on that at the next shorter one.
With only three adjustable parameters this model is able to
reproduce different characteristic properties of intra-day FX
price series.
\\
 Note that the existence of different groups of traders and the
volatility updating mechanism are elements similar to those found in
market microstructure models.
Therefore the SCM may provide a link between market microstructure models
and more conventional models.

One of the authors (S.G.) gratefully acknowledges suppport from the
Swiss National Science Foundation.

\end{document}